\documentclass[11pt]{article}

  \usepackage{latexsym}
  \usepackage{amssymb}

  \def\AFOUR{%
  \setlength{\textheight}{9.0in}%
  \setlength{\textwidth}{5.75in}

  \setlength{\topmargin}{-0.375in}%
  \hoffset=-.5in%
  \renewcommand{\baselinestretch}{1.17}%
  \setlength{\parskip}{6pt plus 2pt}%
  }

  \AFOUR                                          

  \makeatletter
  \def\section{\@startsection {section}{1}{\z@}{-3.5ex plus -1ex minus
   -.2ex}{2.3ex plus .2ex}{\large\bf}}
  \def\subsection{\@startsection{subsection}{2}{\z@}{-3.25ex plus -1ex minus
   -.2ex}{1.5ex plus .2ex}{\normalsize\bf}}
  \makeatother

  \makeatletter
  \@addtoreset{equation}{section}
  
  \makeatother
\newtheorem{theorem}{Theorem}[section]
\newtheorem{proposition}[theorem]{Proposition}

\begin{document}
\begin{titlepage}

\begin{center}

\vspace{1cm}

\vskip .5in
 {\Large\bf Unified $(p,q;\alpha,\beta,\nu;\gamma)-$deformed oscillator algebra: irreducible representations and induced deformed harmonic oscillator}

\vspace{10pt}
E. Baloitcha$^{\dag}$,
 M. N. Hounkonnou$^{\dag}$\footnote{Corresponding author:
norbert.hounkonnou@cipma.uac.bj  with copy to hounkonnou@yahoo.fr}
and E. B. Ngompe Nkouankam$^{\dag}$
\vspace{20pt}

{\em $^\dag$ International Chair in Mathematical Physics
and Applications} \\
{\em (ICMPA-UNESCO Chair)}\\
{\em  University of Abomey-Calavi}\\
{\em  072 B.P. 50 Cotonou, Republic of Benin}\\
\vspace{5pt}

\today

\begin{abstract}
A new  deformed canonical commutation relation, generalizing various known deformations, is defined together with
   its structure function of deformation.  Then,
 the related irreducible representations are characterized and classified. Finally,
    the discrete spectrum of the corresponding deformed harmonic oscillator  Hamiltonian is investigated and discussed.
\end{abstract}
\end{center}

Key-words: Deformed oscillator algebra, algebra representations, discrete spectrum, structure function.

Pacs numbers: 02.20.Uw, 03.65.G, 03.65.Fd

\end{titlepage}
\makeatother

\section{Introduction}

In recent years, a lot of interest has been devoted to the study of  various quantum deformations of bosonic
 oscillators. From the mathematical  point of view,  this  popularity is  due to their connection
 with the non-commutative geometry,
  special functions of multiparameter  analysis \cite{elvis} and other fields of mathematics.  From the other side,
  there are some hopes that  the deformed oscillator can be more suitable for  physical studies of nonlinear phenomena
    than the usual bosonic oscillator of the standard quantum mechanics.  Such hopes
  are supported by several applications in conformal field theory  \cite{witten}, in nuclear  spectroscopy
  \cite{bog,bona}, in nonlinear quantum optics, in condensed matter physics  \cite{Geloun}, and
   in the description of  systems with
  non-standard statistics and energy spectrum  \cite{and}.

 Deformations of the oscillator arose from the successive generalizations of
 the Arik-Coon \cite{arik}  and Biedenharm-Macfarlane \cite{mac1,mac2}
 $q-$oscillators. As in the
 classical case, the problem of realization  of $q-$deformed algebras by the one-parameter deformed creation
 and annihilation operators  is important for the representation theory of quantum groups.
  Various attempts have been
 made to introduce new  parameters in the rich and varied choices of deformed commutation relations.  For instance,
 the study of two-parameter quantum groups, mainly based on the famous   $(p,q)-$ deformation, and their representations
 has started in the works \cite{curt,smir}.
 It is worth noticing that the increase of the number of deformation parameters  makes the method of the
 deformations more flexible.

    In this work, we consider the generalized $(p,q;\alpha,\beta,\nu;\gamma)-$deformed oscillator algebra  as a
 generalization of the $(q;\alpha,\beta,\nu;\gamma)-$deformed oscillator algebra  recently defined in
 \cite{burban},  investigate in detail its representations which are suitably classified, and scrutinize the properties of
 the corresponding deformed harmonic oscillator.

 The paper is organized as follows.
In section 2, we  recall some multiparameter deformations which appear in the literature and define our algebra.
In section 3, we study some relevant properties of this algebra. In section 4, we give a classification of the
 representations of this algebra. We discuss properties of discrete spectrum of the Hamiltonian of the deformed
 harmonic oscillator corresponding to this system and end with some concluding remarks in section 5.

\section{ Multiparameter deformations of the Heisenberg-Weyl algebra: overview and new $(p,q;\alpha,\beta,\nu;\gamma)-$deformed oscillator algebra}

 The nondeformed  oscillator algebra of the quantum harmonic  oscillator is defined by the canonical commutation
 relations
 \begin{eqnarray}
 [a,a^{\dagger}]=I,\qquad [N,a]=-a,\qquad [N,a^{\dagger}]=a^{\dagger}.
 \end{eqnarray}
  The common used deformation of this Heisenberg-Weyl algebra is  defined as the algebra generated by the set of
  operators $\{I, a, a^{\dagger}, N\}$  and  the relations \cite{janus}
  \begin{eqnarray}
 [N,a]=-a,\qquad [N,a^{\dagger}]=a^{\dagger},\qquad a^{\dagger}a=f(N),\qquad aa^{\dagger}=f(N+1)
  \end{eqnarray}
  where  the structure function of the deformation, $f$,    is  positive and  analytic.

  The structure function $f(x)$ characterizes the deformation scheme. In our context,
 let us briefly recall some multiparameter
deformations known in the literature. Other multiparameter generalizations of one-parameter
 deformations were  enumerated in \cite{burban}.

  \begin{itemize}
  \item[$(i)$]
  The  $(p,q)$-generalization of the algebra  introduced by
Chakrabarty and  Jagannathan in \cite{jang}. It is  generated by the operators
 $I$, $a$, $a^{\dagger}$ and  $N$  as follows:
\begin{eqnarray}
aa^{\dagger}-qa^{\dagger}a=p^{-N},\qquad  aa^{\dagger}-p^{-1}a^{\dagger}a = q^{N}  \cr
 [N,a]=-a, \qquad [N,a^{\dagger}]=a^{\dagger}\cr
 f(n)=\frac{p^{-n}-q^{n}}{p^{-1}-q}.
\end{eqnarray}
  \item[$(ii)$]
 The   $(p,q)-$generalization of the
Quesne's deformed bosonic algebra    defined as \cite{ques}:
\begin{eqnarray}
p^{-1}aa^{\dagger}-a^{\dagger}a=q^{-N-1},\qquad qaa^{\dagger}-a^{\dagger}a = p^{N+1}\cr
[N,a]=-a, \qquad [N,a^{\dagger}]=a^{\dagger}\cr
 f(n)=\frac{p^{n}-q^{-n}}{q-p^{-1}}.
\end{eqnarray}
  \item[$(iii)$]
 The $(p,q;\alpha,\beta,l)-$deformed oscillator algebra, given by
the generators $I$, $a$, $a^{\dagger}$, $N$  and the commutation relations \cite{burban1}
\begin{eqnarray}
aa^{\dagger}-q^{l}a^{\dagger}a=p^{-\alpha N-\beta},\qquad  aa^{\dagger}-p^{-l}a^{\dagger}a = q^{\alpha N+\beta}\cr
 [N,a]=-la, \qquad [N,a^{\dagger}]=la^{\dagger}\cr
 f(n)=\frac{p^{-\alpha n-\beta}-q^{\alpha n+\beta}}{p^{-l}-q^{l}}
\end{eqnarray}
with $\alpha, \beta, l\in\mathbb{R}.$
\end{itemize}

In this work, we define the  $(p,q;\alpha,\beta,\nu;\gamma)-$deformed oscillator algebra as:
  \begin{eqnarray}
  \label{alg}
  aa^{\dagger}-p^{\nu}a^{\dagger}a=(1+2\gamma K)q^{\alpha N +\beta},
  \quad [N, a]=-a,\quad [N,a^{\dagger}]=a^{\dagger}\cr
  Ka=-aK,\quad Ka^{\dagger}=-a^{\dagger}K,\quad [N,K]=0,\quad
  N^{\dagger}=N,\quad K^{\dagger}=K
   \end{eqnarray}
 where $p,q\in \mathbb{R}_{+}$, $\alpha,\beta,\nu,\gamma\in \mathbb{R}.$ Particular algebras are readily recovered.
For instance, in the limit $p\to q,$ one finds the   $(q;\alpha,\beta,\nu;\gamma)-$deformed oscillator algebra introduced
 by Burban \cite{burban}. Furthermore, the algebra (\ref{alg}) is also a generalization of the above mentioned $(p,q)-$ algebras.

 \section{Unified $(p,q;\alpha,\beta,\nu;\gamma)-$deformed oscillator algebra: structure function  and pertinent
   properties}

 The generalized
 $(p,q;\alpha,\beta,\nu;\gamma)-$deformed oscillator and its deformed Heisenberg-Weyl algebra are defined by the positive
 structure function $f,$  satisfying $f(0)=0.$ The Fock realization of this algebra covers with
 the form
 \begin{eqnarray}
 a|n\rangle=\sqrt{f(n)}|n-1\rangle,\qquad   a^{\dagger}|n\rangle=\sqrt{f(n+1)}|n+1\rangle\cr
 N|n\rangle= n|n\rangle, \qquad K|n\rangle=(-1)^{n}|n\rangle
 \end{eqnarray}
 with $n\in\mathbb{N}.$

 \begin{proposition}
$\;$ \\
 The structure function  of the  $(p,q;\alpha,\beta,\nu;\gamma)-$deformed oscillator algebra (\ref{alg}) is given by
 \begin{eqnarray}
 f(n)=
\left\{
\begin{array}{cc}

q^{\beta}\left(
\frac{p^{n\nu}-q^{n\alpha}}{p^{\nu}-q^{\alpha}}+
2\gamma\frac{p^{n\nu}-(-1)^{n}q^{n\alpha}}{p^{\nu}+q^{\alpha}}
\right), & \;
{\rm if}\; p^{\nu}\neq q^{\alpha}\cr
 \left(
n + 2\gamma \frac{1-(-1)^{n}}{2}\right)\times q^{(n-1)\alpha +\beta},& \;
{\rm if}\; p^{\nu}=q^{\alpha}.
\end{array}
\right.
\end{eqnarray}
 \end{proposition}

 {\bf Proof.}
  By applying the basis $|n\rangle$ on the relation
   $aa^{\dagger}-p^{\nu}a^{\dagger}a=(1+2\gamma K)q^{\alpha N +\beta}$, we obtain the recurrence relation
 \begin{eqnarray}
 \label{recu}
 f(n+1)-p^{\nu}f(n)=(1+2\gamma (-1)^{n})q^{\alpha n +\beta}
 \end{eqnarray}
 whose  solution is
 \begin{eqnarray}
 \label{rec}
  f(n)=
 \sum_{k=0}^{n-1}
 p^{(n-k-1)\nu}q^{\alpha k+\beta} +2\gamma \sum_{k=0}^{n-1}
 (-1)^{k}p^{(n-k-1)\nu}q^{\alpha k+\beta}.
 \end{eqnarray}
 Indeed, for $n=1$, $f(1)= q^{\alpha \times 0 +\beta}  +2\gamma (-1)^{0}q^{\alpha \times 0 +\beta}.$ Now, we suppose
 that the function $f(k)$ is  given by  (\ref{rec})  for  $k\leq n$. Then we have
 \begin{eqnarray*}
 f(n+1)&=& p^{\nu}f(n) +(1+2\gamma (-1)^{n})q^{\alpha n +\beta}\cr
 &=& p^{\nu}\left(
 \sum_{k=0}^{n-1}
 p^{(n-k-1)\nu}q^{\alpha k+\beta} +2\gamma \sum_{k=0}^{n-1}
 (-1)^{k}p^{(n-k-1)\nu}q^{\alpha k+\beta}
 \right)    \cr
 &&\qquad + (1+2\gamma (-1)^{n})q^{\alpha n +\beta}\cr
 &=&
 \sum_{k=0}^{n}
 p^{(n-k)\nu}q^{\alpha k+\beta} +2\gamma \sum_{k=0}^{n}
 (-1)^{k}p^{(n-k)\nu}q^{\alpha k+\beta}
 \end{eqnarray*}
  which proves the claim.

  Hence, if $p^{\nu}\neq q^{\alpha}$
  \begin{eqnarray*}
 f(n)&=&
p^{(n-1)\nu}q^{\beta}\left(
\frac{1-(p^{-\nu}q^{\alpha})^{n}}{1-p^{-\nu}q^{\alpha}}
+2\gamma
\frac{1-(-1)^{n}(p^{-\nu}q^{\alpha})^{n}}{1+p^{-\nu}q^{\alpha}}
\right)\cr
 &=&
q^{\beta}\left(
\frac{p^{n\nu}-q^{n\alpha}}{p^{\nu}-q^{\alpha}}+
2\gamma\frac{p^{n\nu}-(-1)^{n}q^{n\alpha}}{p^{\nu}+q^{\alpha}}
\right)
\end{eqnarray*}
and
\begin{eqnarray*}
f(n)&=& \left(
\sum_{k=0}^{n-1} 1^{k}+ 2\gamma\sum_{k=0}^{n-1}(-1)^{k}\right)\times q^{(n-1)\alpha +\beta}\cr
&=&
 \left(
n + 2\gamma \frac{1-(-1)^{n}}{2}\right)\times q^{(n-1)\alpha +\beta}
\end{eqnarray*}
if   $p^{\nu} =q^{\alpha}.$
$\Box$

Let us  study  the positivity of the
$(p,q;\alpha,\beta,\nu;\gamma)-$deformed  structure function $f(n)$.
  The inequality $f(n)>0$  can be rewritten as
  \begin{eqnarray}
  \label{even1}
  (p^{n\nu}-q^{n\alpha})\left\{
  \frac{1}{p^{\nu}-q^{\alpha}}+2\gamma\frac{1}{p^{\nu}+q^{\alpha}}\right\} >0
  \end{eqnarray}
  if $n$ is even, and
  \begin{eqnarray}
  \frac{p^{n\nu}-q^{n\alpha}}{p^{n\nu}+q^{n\alpha}}
  \frac{p^{\nu}+q^{\alpha}}{p^{\nu}-q^{\alpha}} + 2\gamma > 0
  \end{eqnarray}
  if $n$ is odd.

 Let us define the sequence
 \begin{eqnarray}
 u_{n}
 =\left\{
 \begin{array}{cc}
  u(\alpha,\nu)\frac{p^{n\nu}-q^{n\alpha}}{p^{n\nu}+q^{n\alpha}},\; &
   {\rm if}\; n>1\\
 1,  \; & {\rm if }\; n=1
 \end{array}
 \right.
 \end{eqnarray}
 where $u(\alpha,\nu)= \frac{p^{\nu}+q^{\alpha}}{p^{\nu}-q^{\alpha}}$.

 If
 \begin{eqnarray*}
 f(x)=  u(\alpha,\nu)\frac{p^{x\nu}-q^{x\alpha}}{p^{x\nu}+q^{x\alpha}}
 \end{eqnarray*}
 we have
 $$\frac{{\rm d}f(x)}{{\rm d}x}=f^{\prime}(x)=
 \frac{(\nu\ln p-\alpha\ln q)}{p^{\nu}-q^{\alpha}}\times \frac{2(p^{\nu}+q^{\alpha})}
 {(p^{x\nu}+q^{x\alpha})^{2}}.$$
  Since $(\nu\ln p-\alpha\ln q)(p^{\nu}-q^{\alpha})>0,$ we deduce that
  $(u_{n})_{n\geq 1}$ is an increasing  sequence. We have $u_{n}>1$ for all $n\in\mathbb{N}\setminus\{0\}$. To have
  $u_{n}+ 2\gamma>0,$ $\forall n\in \mathbb{N}\setminus\{0\},$ it is sufficient that $1+ 2\gamma>0.$

The relation (\ref{even1}) is equivalent  to $-2\gamma<
  \frac{p^{\nu}+q^{\alpha}}{p^{\nu}-q^{\alpha}}$  if $\nu\ln p > \alpha \ln q$ and
$2\gamma<
  -\frac{p^{\nu}+q^{\alpha}}{p^{\nu}-q^{\alpha}}$  if $\nu\ln p < \alpha \ln q.$

Therefore we obtain:
 \begin{proposition}
 The structure function  $f(n)$ is positive for
 \begin{eqnarray}
 -2\gamma<1 \qquad  {\rm if}\qquad  \nu\ln p > \alpha \ln q
 \end{eqnarray}
 and
 \begin{eqnarray}
 -1<2\gamma<- \frac{p^{\nu}+q^{\alpha}}{p^{\nu}-q^{\alpha}}
 \qquad  {\rm if}\qquad \nu\ln p < \alpha \ln q.
 \end{eqnarray}
\end{proposition}

 \begin{proposition}
  From the relation $aa^{\dagger}-p^{\nu}a^{\dagger}a=(1+2\gamma K)q^{\alpha N +\beta},$ we get
   the formula
     \begin{eqnarray}
     \label{com}
    a(a^{\dagger})^{n}-
 p^{n\nu}(a^{\dagger})^{n}a
 =\left[
 n;\alpha,\beta,\nu;\gamma K
 \right](a^{\dagger})^{n-1}q^{\alpha N +\beta}, \quad \forall n\in\mathbb{N}\setminus\{0\}
 \end{eqnarray}
where
\begin{eqnarray*}
\label{com1}
  \left[
 n;\alpha,\beta,\nu;\gamma K
 \right]=
 \left\{
 \begin{array}{cc}
\frac{p^{n\nu}-q^{n\alpha}}{p^{\nu}-q^{\alpha}} + 2\gamma K
  \frac{q^{n\alpha}-(-1)^{n}p^{n\nu}}{p^{\nu}+q^{\alpha}},
   & \;{\rm if }\;p^{\nu}\neq q^{\alpha}\\
 nq^{\alpha(n-1)} + 2\gamma K q^{\alpha(n-1)}\times \left(
 \frac{1-(-1)^{n}}{2}
 \right), &\;{\rm if }\;p^{\nu}= q^{\alpha}.
 \end{array}
 \right.
\end{eqnarray*}

\end{proposition}

{Proof.}

Let us prove  (\ref{com}) by recurrence. It is obviously true for $n=1$. For $n=2$,  we have
\begin{eqnarray*}
  a(a^{\dagger})^{2}&=&aa^{\dagger}a^{\dagger}\cr
 &=& (p^{\nu}a^{\dagger}a+(1+2\gamma K )q^{\alpha N +\beta})a^{\dagger}\cr
  &=& p^{2\nu}(a^{\dagger})^{2}a + p^{\nu}a^{\dagger}(1+2\gamma K )q^{\alpha N +\beta}
 + (1+2\gamma K )q^{\alpha N +\beta}a^{\dagger}\cr
 \ &=& p^{2\nu}(a^{\dagger})^{2}a + p^{\nu}(1-2\gamma K )a^{\dagger}q^{\alpha N +\beta}
 + (1+2\gamma K )a^{\dagger}q^{\alpha( N+1) +\beta}
\end{eqnarray*}
so that
\begin{eqnarray*}
 a(a^{\dagger})^{2}-
 p^{2\nu}(a^{\dagger})^{2}a =\left\{
p^{\nu} +q^{\alpha} +  2\gamma K(q^{\alpha}-p^{\nu})  \right\} a^{\dagger}q^{\alpha N +\beta}.
\end{eqnarray*}
This latter relation can be put in the form
\begin{eqnarray}
 a(a^{\dagger})^{2}-
 p^{2\nu}(a^{\dagger})^{2}a =\left\{
 \frac{p^{2\nu}-q^{2\alpha}}{p^{\nu}-q^{\alpha}} + 2\gamma K
  \frac{q^{2\alpha}-(-1)^{2}p^{2\nu}}{p^{\nu}+q^{\alpha}}
 \right\}a^{\dagger}q^{\alpha N +\beta}
 \end{eqnarray}
 if $p^{\nu}\neq q^{\alpha},$  and
 \begin{eqnarray}
 a(a^{\dagger})^{2}-
 p^{2\nu}(a^{\dagger})^{2}a =\left\{
 2q^{\alpha} + 2\gamma K q^{\alpha(2-1)}\times \left(
 \frac{1-(-1)^{2}}{2}
 \right)
 \right\}a^{\dagger}q^{\alpha N +\beta}
 \end{eqnarray}
if $p^{\nu}=q^{\alpha}$.

We claim that  the relation (\ref{com}) is true for $k\leq n$. Then, we have
\begin{eqnarray*}
 a(a^{\dagger})^{n+1}
 &=& a(a^{\dagger})^{n}a^{\dagger}\cr
 &=& \left\{
p^{n\nu}(a^{\dagger})^{n}a
 +\left[
 n;\alpha,\beta,\nu;\gamma K
 \right](a^{\dagger})^{n-1}q^{\alpha N +\beta}
\right\} a^{\dagger}\cr
 &=&
p^{n\nu}(a^{\dagger})^{n}
\left\{
p^{\nu}a^{\dagger}a+(1+2\gamma K)q^{\alpha N +\beta}
\right\} \cr
&&\qquad +
\left[
 n;\alpha,\beta,\nu;\gamma K
 \right](a^{\dagger})^{n}q^{\alpha(N+1) +\beta} \cr
 &=&
p^{(n+1)\nu}(a^{\dagger})^{n+1}a
+p^{n\nu}(1+2(-1)^{n}\gamma K)(a^{\dagger})^{n}q^{\alpha N +\beta}\cr
&& \qquad  + q^{\alpha}
\left[
 n;\alpha,\beta,\nu;\gamma K
 \right](a^{\dagger})^{n}q^{\alpha N +\beta}.
 \end{eqnarray*}
 So,
 \begin{eqnarray}
&&
  a(a^{\dagger})^{n+1}-p^{(n+1)\nu}
(a^{\dagger})^{n+1}a
= \left\{
p^{n\nu} + q^{\alpha}\left(
\frac{p^{n\nu}-q^{n\alpha}}{p^{\nu}-q^{\alpha}}\right)\right.\cr
&&\quad \left. +2\gamma K \left[
(-1)^{n}p^{n\nu} + q^{\alpha}\frac{q^{n\alpha}-(-1)^{n}p^{n\nu}}{p^{\nu}+q^{\alpha}}
\right]
\right\}
(a^{\dagger})^{n}q^{\alpha N +\beta} \cr
&=&
\left[
 n+1;\alpha,\beta,\nu;\gamma K
 \right](a^{\dagger})^{n}q^{\alpha N +\beta}
  \end{eqnarray}
  if $p^{\nu}\neq q^{\alpha}$ and
  \begin{eqnarray}
 && a(a^{\dagger})^{n+1}-p^{(n+1)\nu}
(a^{\dagger})^{n+1}a \cr
&=& \left\{q^{n\alpha }(1+2(-1)^{n}\gamma K) +
q^{\alpha}\left[
 n;\alpha,\beta,\nu;\gamma K
 \right]
 \right\}(a^{\dagger})^{n}q^{\alpha N +\beta} \cr
 &=&
 \left\{(n+1)q^{n\alpha } + 2\gamma K q^{n\alpha }\times \left(
 \frac{1-(-1)^{n+1}}{2}
 \right)\right\}(a^{\dagger})^{n}q^{\alpha N +\beta} \cr
 &=&\left[
 n+1;\alpha,\beta,\nu;\gamma K
 \right](a^{\dagger})^{n}q^{\alpha N +\beta}
 \end{eqnarray}
  if $p^{\nu}= q^{\alpha}$,   which proves the claim.
$\Box$

We readily obtain the generated  function for $\left[
 n;\alpha,\beta,\nu;\gamma K
 \right]:$

 \begin{eqnarray}
 \sum_{n=0}^{+\infty}\left[
 n;\alpha,\beta,\nu;\gamma K
 \right]z^{n}
 =\left\{
 \begin{array}{cc}
 \frac{z}{1-q^{\alpha}z}
\left(\frac{1}{1-p^{\nu}z} + 2\gamma K \frac{1}{1+p^{\nu}z}\right),
& \; {\rm if }\; p^{\nu}\neq q^{\alpha} \\
z\left(\frac{1}{(1-q^{\alpha}z)^{2}} + 2\gamma K \frac{1}{1-q^{2\alpha}z^{2}}\right),
& \; {\rm if }\; p^{\nu}= q^{\alpha}.
 \end{array}
 \right.
 \end{eqnarray}

  \section{Irreducible representations of the unified
$(p,q;\alpha,\beta,\nu;\gamma)-$deformed oscillator: characterization and classification}

Here we give, following the work by Rideau \cite{rideau}, the classification of the
irreducible representations of the  $(p,q;\alpha,\beta,\nu;\gamma)-$deformed oscillator  algebra.

This algebra possesses the following Casimir operators
\begin{eqnarray}
C_{1}=K^{2},\qquad C_{2}=Ke^{i\pi N}, \qquad C_{3}=e^{2 i\pi N}.
\end{eqnarray}
Let $w$ be an eigenvalue of $C_{2}$ corresponding to a given representation. Then,
$K=we^{-i\pi N}$.
Let $\psi_{0}$ be  a common eigenvector of $N$  and $K$:
\begin{eqnarray}
N\psi_{0}&=&\nu_{0}\psi_{0}\\
K\psi_{0}&=& \gamma e^{-i\nu_{0}\pi}\psi_{0}.
\end{eqnarray}
Due  to  the commutativity of  $a^{\dagger}a$ and  $aa^{\dagger}$  with $N$ and $K$, we may assume that:
\begin{eqnarray}
aa^{\dagger}\psi_{0}&=& \lambda_{0}\psi_{0} \\
a^{\dagger}a\psi_{0}&=&\mu_{0}\psi_{0}
\end{eqnarray}
and $(\psi_{0},\psi_{0})=1.$   One can check that the vectors defined by
\begin{eqnarray}
\phi_{n}=
\left\{
\begin{array}{cc}
(a^{\dagger})^{n}\psi_{0},&\; {\rm if }\; n\geq 0\\
 a^{-n}\psi_{0},&\; {\rm if }\; n< 0
\end{array}
\right.
\end{eqnarray}
are eigenvectors of $a^{\dagger}a$ and $aa^{\dagger}:$
\begin{eqnarray}
a^{\dagger}a\phi_{n}&=& \lambda_{n}\phi_{n} \\
 aa^{\dagger}\phi_{n}&=&\mu_{n}\phi_{n}.
\end{eqnarray}

Now, let us define the following vectors:
\begin{eqnarray}
\psi_{n}
=\left\{
\begin{array}{cc}
\frac{1}{\sqrt{\prod_{k=1}^{n}\lambda_{k}}} (a^{\dagger})^{n}\psi_{0},& \; {\rm for }\; n\geq 0\\
 \frac{1}{\sqrt{\prod_{k=1}^{-n}\lambda_{k+n}}} a^{-n}\psi_{0},& \; {\rm for }\; n< 0
\end{array}
\right.
\end{eqnarray}
which are orthogonal states. The actions of  basic operators on them are given by

\begin{eqnarray}
a^{\dagger}\psi_{n}&=&\sqrt{\lambda_{n+1}}\psi_{n+1}\label{rep1}\\
a\psi_{n}&=&\sqrt{\lambda_{n}}\psi_{n-1}\label{rep2}\\
N\psi_{n}&=&(\nu_{0}+n)\psi_{n}\label{rep3}\\
K\psi_{n}&=&\frac{(-1)^{n}}{2\gamma}B \psi_{n} \label{rep4}
\end{eqnarray}

 where $B=2\gamma w e^{-i\pi \nu_{0}}\in \mathbb{R}.$

 The additional condition that we have to take into account is that $\lambda_{n}$ and $\mu_{n-1},$  being
 eigenvalues of nonnegative operators $a^{\dagger}a$  and $aa^{\dagger}$, respectively, should be nonnegative.

 Applying   the relation
  $aa^{\dagger}-p^{\nu}a^{\dagger}a=(1+2\gamma K)q^{\alpha N +\beta}$ on the vectors $\psi_{n}$ yields
  $\lambda_{n+1}=p^{\nu}\lambda_{n} + q^{\alpha \nu_{0}+\beta}(1+(-1)^{n}B)q^{\alpha n}$ and
one can prove by recurrence that
\begin{eqnarray}
\lambda_{n}
=\left\{
\begin{array}{cc}
 p^{n\nu}\lambda_{0}
  + q^{\alpha\nu_{0}+ \beta}
\times
\left(
  \frac{p^{n\nu }-q^{n\alpha }}{ p^{\nu}-q^{\alpha}}
  + B \frac{p^{n\nu }-(-1)^{n}q^{n\alpha}}{ p^{\nu}+q^{\alpha}} \right),
&\; {\rm if }\; p^{\nu}\neq q^{\alpha} \\
q^{n\alpha}\lambda_{0}
  + q^{\alpha\nu_{0}+ \beta}
  \times q^{(n-1)\alpha} \left[
  n
  + B \left(\frac{1-(-1)^{n}}{2}\right)
  \right],
&\; {\rm if }\; p^{\nu}= q^{\alpha}.
\end{array}
\right.
\end{eqnarray}

\begin{itemize}
\item[$(A)$]
If $p^{\nu}= q^{\alpha}$, the condition $\lambda_{n}\geq 0$  is equivalent to
\begin{eqnarray}
\lambda_{0}\geq
-q^{\alpha (\nu_{0}-1)+\beta}\left(n + B\frac{1-(-1)^{n}}{2}\right).
\end{eqnarray}
Since
\begin{eqnarray}
\lim_{n\to -\infty}
 -q^{\alpha (\nu_{0}-1)+\beta}\left(n + B\frac{1-(-1)^{n}}{2}\right) =+\infty,
\end{eqnarray}
there exists an integer $n_0$  for which $\lambda_{n}\leq 0$  for $n\leq n_{0}.$ Since $a^{\dagger}a \geq 0,
$  we have
\begin{eqnarray}
a\psi_{n}=0 \quad {\rm for }\quad n\leq n_{0}.
\end{eqnarray}
After possible renumbering, we may assume
\begin{eqnarray}
a\psi_{0}=0,\qquad \lambda_{0}=0.
\end{eqnarray}
Therefore, the representation is spanned by the vectors $\psi_{n},$ $n\geq 0$ and the eigenvalues $\lambda_{n}$ are given
by
\begin{eqnarray}
\lambda_{n}=
n q^{\alpha(\nu_{0}+n-1)+ \beta}
  + B q^{\alpha(\nu_{0}+n-1)+ \beta}\times\left(\frac{1-(-1)^{n}}{2}\right).
\end{eqnarray}
The arbitrary values of the parameters $\nu_{0}$ and $B\geq 0$ correspond to non-equivalent representations of
(\ref{rep1})-(\ref{rep4}).
\item[$(B)$]
If $p^{\nu}\neq q^{\alpha}$, $\lambda_{n}$  can be rewritten as
\begin{eqnarray}
\label{eigen1}
 \lambda_{n}
&=&p^{n\nu}\left\{
\lambda_{0}  + q^{\alpha\nu_{0}+\beta}
\left[
\frac{1}{p^{\nu}-q^{\alpha}} + \frac{B}{p^{\nu}+q^{\alpha}}\right.\right. \cr
&&\qquad \left.\left. -(p^{-\nu}q^{\alpha})^{n}\left(
\frac{1}{p^{\nu}-q^{\alpha}} + \frac{B(-1)^{n}}{p^{\nu}+q^{\alpha}}
\right)
\right]
\right\}
\end{eqnarray}
 so that  its  positivity  is equivalent to
 \begin{eqnarray*}
\lambda_{0} q^{-(\alpha\nu_{0}+\beta)}
+ \frac{1}{p^{\nu}-q^{\alpha}} + \frac{B}{p^{\nu}+q^{\alpha}}
 \geq(p^{-\nu}q^{\alpha})^{2k}\left(
\frac{1}{p^{\nu}-q^{\alpha}} + \frac{B}{p^{\nu}+q^{\alpha}}
\right)
 \end{eqnarray*}
 and

 \begin{eqnarray*}
  \lambda_{0} q^{-(\alpha\nu_{0}+\beta)}
+ \frac{1}{p^{\nu}-q^{\alpha}} + \frac{B}{p^{\nu}+q^{\alpha}}
\geq(p^{-\nu}q^{\alpha})^{2k+1}\left(
\frac{1}{p^{\nu}-q^{\alpha}} -\frac{B}{p^{\nu}+q^{\alpha}}
\right).
 \end{eqnarray*}

\item[$(B)_{1}$]
 Assume that $\nu\ln p > \alpha \ln q$. Then, at least one of
 $\frac{1}{p^{\nu} - q^{\alpha}} \pm \frac{B}{p^{\nu} + q^{\alpha}}$ is positive.
 Therefore, there exists an integer  $n_{0}$  such that for $n$ odd/ or even
 $\lambda_{n}\leq 0$  for $n\leq n_{0},$  which implies $a\psi_{n}=0$ for some
 $n\leq n_{0}$. After possible renumbering, we may assume
 \begin{eqnarray}
 a\psi_{0}=0,\qquad \lambda_{0}=0.
 \end{eqnarray}
 Therefore, the representations are given by
 \begin{eqnarray}
 \label{eigen3}
\lambda_{n} = q^{\alpha\nu_{0}+ \beta}
\times
\left(
  \frac{p^{n\nu }-q^{n\alpha }}{ p^{\nu}-q^{\alpha}}
  + B \frac{p^{n\nu }-(-1)^{n}q^{n\alpha }}{ p^{\nu}+q^{\alpha}} \right).
 \end{eqnarray}

Let us now study the positivity of $\lambda_{n}$.

 If $n$  is even,
 \begin{eqnarray}
 \lambda_{n}
 = q^{\alpha\nu_{0}+ \beta}
(p^{n\nu n}-q^{n\alpha})
\left(
  \frac{1}{ p^{\nu}-q^{\alpha}}
  + \frac{B}{ p^{\nu}+q^{\alpha}} \right)
 \end{eqnarray}
and $\lambda_{n}$ is positive when $B\geq -\frac{p^{\nu}+q^{\alpha}}{p^{\nu}-q^{\alpha}}.$

If $n$ is odd, $\lambda_{n}$ is positive if and only if
\begin{eqnarray}
\label{ineq1}
 \frac{p^{\nu n}-q^{\alpha n}}{ p^{n\nu}+q^{n\alpha}}
 \frac{p^{\nu }+q^{\alpha }}{ p^{\nu}-q^{\alpha}} + B \geq 0
 \end{eqnarray}
 which  is satisfied when $B\geq -1.$

 We conclude that $\lambda_{n}$  is positive for all $n\in \mathbb{N}$
 if $B\geq -1$ since  $-\frac{p^{\nu}+q^{\alpha}}{p^{\nu}-q^{\alpha}}<-1$.

 For $B=-1,$ $\lambda_{1}=0.$ Therefore, we obtain
 \begin{eqnarray}
 \label{one}
 a=a^{\dagger}=0;\qquad N=\nu_{0};\qquad K=-\frac{1}{2\gamma}.
 \end{eqnarray}
 This representation is one-dimensional. For $B>-1$, the representation is spanned by the vectors
  $\psi_{n}$, $n\geq 0.$ It is defined by (\ref{rep1})-(\ref{rep4}) and (\ref{eigen3}) with $n\geq 0.$

  \item[$(B)_{2}$]
  Assume that $\nu\ln p < \alpha \ln q$ and one of the values
  $\frac{1}{p^{\nu}-q^{\alpha}} \pm \frac{B}{p^{\nu}+q^{\alpha}}$ is positive. In this case, there exists
   an integer $n_{0}$ such that for $n\geq n_{0}$, $\lambda_{n}$
    is negative for even or odd $n_{0}$.
   This implies $a^{\dagger}\psi_{n}=0$ for some $n\geq n_{0}.$  After possible renumbering, we get
   $a^{\dagger}\psi_{0}=0.$ From the relation $aa^{\dagger}\psi_{0}=\mu_{0}\psi_{0}$, we obtain $\mu_{0}=0.$
   Hence, $\lambda_{0}=-p^{-\nu}q^{\alpha\nu_{0}+\beta}(1+B).$

   The condition $\lambda_{0}\geq 0$ is equivalent to $B\leq -1.$ It leads to
   \begin{eqnarray}
   \label{eigen4}
   \lambda_{n}
   &=& q^{\alpha\nu_{0}+\beta}\left\{
   -p^{(n-1)\nu}(1+B) + \frac{p^{n\nu}-q^{n \alpha}}{p^{\nu}-q^{\alpha}}\right.\cr
    && \qquad \qquad \left. +  B \frac{p^{n\nu}-(-1)^{n}q^{n \alpha}}{p^{\nu}+q^{\alpha}}
   \right\}
   \end{eqnarray}
   for $n\in\mathbb{Z}_{-}$.

   Let us now discuss the positivity of $\lambda_{n}$.

   If $n$ is odd, (\ref{eigen4}) can be rewritten as
   \begin{eqnarray}
   \lambda_{n}= q^{\alpha\nu_{0}+\beta}q^{n \alpha }
   (1-(q^{-\alpha}p^{\nu})^{n-1})\left(\frac{1}{ p^{\nu}-q^{\alpha}}
  -  \frac{B}{ p^{\nu}+q^{\alpha}}\right)
   \end{eqnarray}
   which is positive when $B\leq \frac{p^{\nu}+q^{\alpha}}{p^{\nu}-q^{\alpha}}.$

   Now, we suppose that $B\leq \frac{p^{\nu}+q^{\alpha}}{p^{\nu}-q^{\alpha}}.$ When $n$ is even,
   we have
   \begin{eqnarray}
   \lambda_{n}&=&
   q^{\alpha\nu_{0}+\beta}
   \left\{
 q^{\alpha}p^{(n-1)\nu}
 \left(\frac{1}{ p^{\nu}-q^{\alpha}}
  -  \frac{B}{ p^{\nu}+q^{\alpha}}\right) \right.\cr
    && \qquad \qquad \left.-q^{n\alpha}
  \left(\frac{1}{ p^{\nu}-q^{\alpha}}
  + \frac{B}{ p^{\nu}+q^{\alpha}}\right)
   \right\}
   \end{eqnarray}
   which is always positive since $ B< - \frac{p^{\nu}+q^{\alpha}}{p^{\nu}-q^{\alpha}}$.

   We deduce that $\lambda_{n}$ is positive  for all $n\in \mathbb{Z}_{-}$  if
   $B\leq \frac{p^{\nu}+q^{\alpha}}{p^{\nu}-q^{\alpha}}$.

   For $B< \frac{p^{\nu}+q^{\alpha}}{p^{\nu}-q^{\alpha}}$, the representation is  given by
   (\ref{rep1})-(\ref{rep4}) and (\ref{eigen4}) for $n\leq 0.$

   For $B=\frac{p^{\nu}+q^{\alpha}}{p^{\nu}-q^{\alpha}}$, we have $\lambda_{n}=0$ if $n$ is  odd
   and $\lambda_{n}=\frac{2 q^{\alpha(\nu_{0}+n)+\beta}}{q^{\alpha}-p^{\nu}}$ if $n$ is even. The representation
   is two-dimensional:
   \begin{eqnarray}
   \begin{array}{cc}
      a\psi_{0}=\sqrt{\frac{2 q^{\alpha\nu_{0}+\beta}}{q^{\alpha}-p^{\nu}}}\psi_{-1};&
   \qquad a\psi_{-1}=0 \label{two1}\\
   a^{\dagger}\psi_{0}=0;&\qquad
   a^{\dagger}
   \psi_{-1}=\sqrt{\frac{2 q^{\alpha\nu_{0}+\beta}}{q^{\alpha}-p^{\nu}}}\psi_{0} \label{two2} \\
   N\psi_{0}=\nu_{0}\psi_{0};& \qquad   N\psi_{-1}=(\nu_{0}-1)\psi_{-1}\label{two3}\\
   K\psi_{0}=\frac{1}{2\gamma}\frac{p^{\nu}+q^{\alpha}}{p^{\nu}-q^{\alpha}}\psi_{0};&
   \qquad
   K\psi_{-1}=-\frac{1}{2\gamma}\frac{p^{\nu}+q^{\alpha}}{p^{\nu}-q^{\alpha}}\psi_{-1}. \label{two4}
   \end{array}
   \end{eqnarray}

  \item[$(B)_{3}$]
   Assume $\nu\ln p<\alpha \ln q$ and both values $\frac{1}{p^{\nu}-q^{\alpha}}\pm
  \frac{B}{p^{\nu}+q^{\alpha}}$ are non positive. Let us consider the following cases:
  \begin{itemize}

  \item[${(a)}:$]
  \begin{center}
  $\lambda q^{-(\alpha\nu_{0}+\beta)}
  + \frac{1}{p^{\nu}-q^{\alpha}}+
  \frac{B}{p^{\nu}+q^{\alpha}} >0$
  \end{center}

  We have $\lambda_{n}>0,$  for all $n\in\mathbb{Z}$ and the representation is given by
  (\ref{rep1})-(\ref{rep4}) and (\ref{eigen1}) with $n\in\mathbb{Z}$.

  \item[${(b)}:$]
  \begin{center}
   $\lambda q^{-(\alpha\nu_{0}+\beta)}
  + \frac{1}{p^{\nu}-q^{\alpha}}+
  \frac{B}{p^{\nu}+q^{\alpha}} =0 $
  \end{center}

  In this case the condition $\lambda_{n}\geq 0$  for all $n\in \mathbb{Z}$  is equivalent to
  $|B|\leq -\frac{p^{\nu}+q^{\alpha}}{p^{\nu}-q^{\alpha}}.$

  If $|B|< -\frac{p^{\nu}+q^{\alpha}}{p^{\nu}-q^{\alpha}},$ the representation is given by    (\ref{rep1})-
  (\ref{rep4}), where
  \begin{eqnarray}
  \lambda_{n}=-q^{\alpha\nu_{0}+\beta}q^{n\alpha }\left(  \frac{1}{p^{\nu}-q^{\alpha}}+
  \frac{B(-1)^{n}}{p^{\nu}+q^{\alpha}}
  \right)
  \end{eqnarray}
  for all $n\in \mathbb{Z}.$

  If $B = -\frac{p^{\nu}+q^{\alpha}}{p^{\nu}-q^{\alpha}}$, $\lambda_{n}=0$ when $n$ is even. Therefore,
  the representation  is given by:

   \begin{eqnarray}
   \begin{array}{cc}
   a^{\dagger}\psi_{0}=\sqrt{\frac{2 q^{\alpha(\nu_{0}+1)+\beta}}{q^{\alpha}-p^{\nu}}}\psi_{1};&
   \qquad a^{\dagger}\psi_{1}=0 \\
   a\psi_{0}=0;&\qquad
   a\psi_{1}=\sqrt{\frac{2 q^{\alpha(\nu_{0}+1)+\beta}}{q^{\alpha}-p^{\nu}}}\psi_{0}  \\
   N\psi_{0}=\nu_{0}\psi_{0};&\qquad   N\psi_{1}=(\nu_{0}+1)\psi_{1}\\
K\psi_{0}=\frac{1}{2\gamma}\frac{p^{\nu}+q^{\alpha}}{q^{\alpha}-p^{\nu}}\psi_{0};&
   \qquad
   K\psi_{1}=\frac{1}{2\gamma}\frac{p^{\nu}+q^{\alpha}}{p^{\nu}-q^{\alpha}}\psi_{1}.
   \end{array}
   \end{eqnarray}

 If $B = \frac{p^{\nu}+q^{\alpha}}{p^{\nu}-q^{\alpha}}$, $\lambda_{n}=0$ when $n$ is odd and
  $\lambda_{n}= \frac{2 q^{\alpha(\nu_{0}+n)+\beta}}{q^{\alpha}-p^{\nu}}$ when $n$ is even. Therefore,
  the representation is two-dimensional  and is given by (\ref{two1}).

  \item[${(c)}:$]
  \begin{center}
   $\lambda q^{-(\alpha\nu_{0}+\beta)}
  + \frac{1}{p^{\nu}-q^{\alpha}}+
  \frac{B}{p^{\nu}+q^{\alpha}} <0$
 \end{center}
  There exists $n_{0}$  such that $\lambda_{n}\leq 0$ for $n\leq n_{0},$ n even and odd.
  Therefore, the representation
  is given by (\ref{rep1})-(\ref{rep4}) and (\ref{eigen3}) with $n\in\mathbb{N}.$ To provide $\lambda_{n}\geq 0$
  for  all $n\geq 0$ we have to restrict $B$ in the interval
  \begin{eqnarray}
  -1\leq B< -\frac{p^{\nu}+q^{\alpha}}{p^{\nu}-q^{\alpha}}.
  \end{eqnarray}
  For $B=-1$, the representation is given by (\ref{one}).
  \end{itemize}

\end{itemize}

\section{$(p,q;\alpha,\beta,\nu;\gamma)-$deformed oscillator: Hamiltonian definition and energy spectrum computation}

The  Hamiltonian  of the $(p,q;\alpha,\beta,\nu;\gamma)-$deformed oscillator, defined in an analogous way as that of the usual harmonic oscillator as:
\begin{eqnarray}
{\cal H} =\frac{\hbar w}{2}(aa^{\dagger}+a^{\dagger}a),
\end{eqnarray}
 can be rewritten in terms of the  number operator $N$  as:
\begin{eqnarray}
{\cal H} &=&\frac{\hbar w}{2}(f(N)+f(N+1))\cr
&=&\frac{\hbar w}{2}q^{\beta}\left\{
\frac{p^{\nu N} -q^{\alpha N}}{p^{\nu} -q^{\alpha}} + 2\gamma
\frac{p^{\nu N} -(-1)^{N}q^{\alpha N}}{p^{\nu} +q^{\alpha}}\right.\cr
&&\quad\
+ \left.
\frac{p^{\nu(N+1)} -q^{\alpha(N+1)}}{p^{\nu} -q^{\alpha}} + 2\gamma
\frac{p^{\nu (N+1)} -(-1)^{N+1}q^{\alpha(N+1)}}{p^{\nu} +q^{\alpha}}
\right\}.
\end{eqnarray}
In the basis $\{|n\rangle\}$, it has the diagonal form
\begin{eqnarray}
{\cal H}  |n\rangle= e_{n}|n\rangle.
\end{eqnarray}
The energy spectrum, $e_{n}$,  is given by
\begin{eqnarray}
e_{n}&=&\frac{\hbar w}{2}q^{\beta}\left\{
\frac{p^{\nu n} -q^{\alpha n}}{p^{\nu} -q^{\alpha}} + 2\gamma
\frac{p^{\nu n} -(-1)^{n}q^{\alpha n}}{p^{\nu} +q^{\alpha}}\right.\cr
&&\quad
+ \left.
\frac{p^{\nu(n+1)} -q^{\alpha(n+1)}}{p^{\nu} -q^{\alpha}} + 2\gamma
\frac{p^{\nu (n+1)} -(-1)^{n+1}q^{\alpha(n+1)}}{p^{\nu} +q^{\alpha}}
\right\}.
\end{eqnarray}
As a matter of spectra interpretation, let us  introduce a new parametrization:
\begin{eqnarray}
p=\exp(\tau);\quad q=\exp(\rho); \quad \tau\nu=k+\mu;\quad\rho\alpha=k-\mu.
\end{eqnarray}
We  readily obtain
\begin{eqnarray*}
  e_{n}&=& \frac{\hbar w}{2}e^{\rho\beta+k n}
\left\{
\frac{\sinh(\mu(n+1))}{\sinh(\mu)}
 + e^{-k}\frac{\sinh(\mu n)}{\sinh(\mu)}\right.\cr
  && \qquad\qquad \left.
+ 2\gamma\frac{1-(-1)^{n}}{2}
\left(
\frac{\sinh(\mu(n+1))}{\cosh(\mu)}
+ e^{-k}\frac{\cosh(\mu n)}{\cosh(\mu)}\right)
\right.
\cr
  && \qquad\qquad
\left. + 2\gamma\frac{1+
(-1)^{n}}{2}
\left(
\frac{\cosh(\mu(n+1))}{\cosh(\mu)}
+ e^{-k}\frac{\sinh(\mu n)}{\cosh(\mu)}\right)
\right\}
\end{eqnarray*}
or
\begin{eqnarray*}
 e_{n}
&=&
\frac{\hbar w}{2}e^{\rho\beta-k}
\left\{
\frac{1+e^{k+\mu}}{2}\frac{e^{(k+\mu)n}}{\sinh(\mu)}
-\frac{1+e^{k-\mu}}{2}\frac{e^{(k-\mu)n}}{\sinh(\mu)}
\right.\cr
&& \qquad \qquad + \left.
2\gamma\frac{1-(-1)^{n}}{2}
\left(
\frac{1+e^{k+\mu}}{2}\frac{e^{(k+\mu)n}}{\cosh(\mu)}
+ \frac{1-e^{k-\mu}}{2}\frac{e^{(k-\mu)n}}{\cosh(\mu)}
\right)
\right. \cr
&&
\left.
  \qquad \qquad + 2\gamma\frac{1+(-1)^{n}}{2}
\left(
\frac{1+e^{k+\mu}}{2}\frac{e^{(k+\mu)n}}{\cosh(\mu)}
-\frac{1-e^{k-\mu}}{2}\frac{e^{(k-\mu)n}}{\cosh(\mu)}
\right)
\right\}.
\end{eqnarray*}
It is convenient to  separately consider the eigenvalue $e_{n}$ of ${\cal H}$ for $n$ even and odd:
\begin{eqnarray}
e_{n}
&=&
\frac{\hbar w}{2}e^{\rho\beta-k}
\left\{
\frac{1+e^{k+\mu}}{2}\frac{e^{(k+\mu)n}}{\sinh(\mu)}
-\frac{1+e^{k-\mu}}{2}\frac{e^{(k-\mu)n}}{\sinh(\mu)}
\right.\cr
&&   \qquad\qquad +\left.
2\gamma
\left(
\frac{1+e^{k+\mu}}{2}\frac{e^{(k+\mu)n}}{\cosh(\mu)}
+ \frac{1-e^{k-\mu}}{2}\frac{e^{(k-\mu)n}}{\cosh(\mu)}
\right)
\right\}
\end{eqnarray}
for $n$ odd, and
 \begin{eqnarray}
e_{n}
&=&
\frac{\hbar w}{2}e^{\rho\beta-k}
\left\{
\frac{1+e^{k+\mu}}{2}\frac{e^{(k+\mu)n}}{\sinh(\mu)}
-\frac{1+e^{k-\mu}}{2}\frac{e^{(k-\mu)n}}{\sinh(\mu)}
\right.\cr
&&   \qquad\qquad+ \left.
2\gamma
\left(
\frac{1+e^{k+\mu}}{2}\frac{e^{(k+\mu)n}}{\cosh(\mu)}
- \frac{1-e^{k-\mu}}{2}\frac{e^{(k-\mu)n}}{\cosh(\mu)}
\right)
\right\}
\end{eqnarray}
for $n$ even.
From  this result,  the spectrum of this Hamiltonian is not equidistant and the spacing is given by
\begin{eqnarray}
 e_{2n+1}-e_{2n}
&=&\frac{\hbar w}{2}e^{\rho\beta}e^{(2n-1)k}
\left(
\frac{1}{\sinh(\mu)} + 2\gamma\frac{1}{\cosh(\mu)}
\right)\cr
&& \qquad \qquad \times(e^{2k}\sinh(2(n+1)\mu)-\sinh(2 n\mu)).
\end{eqnarray}


According to the above analysis,  the spectrum of the
Hamiltonian is defined for the parameter $-1<2\gamma$ if $\mu>0$ and $-1<2\gamma<-\coth(\mu)$ if $\mu<0.$
In the special case $\mu=0$, it is reduced to
\begin{eqnarray}
e_{n}=
\frac{\hbar w}{2}e^{\rho\beta+k n}
\left\{
(n+\gamma)(1+ e^{-k}) + \gamma(-1)^{n}(1-e^{-k}) + 1
\right\}.
\end{eqnarray}
If, additionally, $k=0$, $\gamma= 0,$ and
 $$(\rho= 0,\quad\mbox{or}\quad\beta=0,\quad\mbox{or}\quad\beta\quad(\mbox{resp.}\quad\rho)\rightarrow -\infty\quad\mbox{for finite}\quad \rho \quad(\mbox{resp.}\quad \beta)),$$   then, we recover the spectrum of the ordinary one-dimensional harmonic oscillator:
\begin{eqnarray}
e_{n}= \hbar w \left (n+{1 \over 2}\right),
\end{eqnarray}
 as expected.

\section{Concluding remarks}
In this work, we have defined a
$(p,q;\alpha,\beta,\nu;\gamma)-$deformed oscillator  algebra which is a  straightforward
generalization of  known deformed algebras. In particular, the Burban multiparameter deformed algebra studied in \cite{burban} has been recovered. The deformation structure function
 and
relevant useful formulas have been computed. Then, the corresponding irreducible representations  have been classified. Finally,
 the spectrum of the deformed oscillator Hamiltonian has been investigated and discussed.

\section*{Acknowledgments}
This work is partially
supported by the Abdus Salam International
Centre for Theoretical
Physics (ICTP, Trieste, Italy) through the
Office of External
Activities (OEA) - \mbox{Prj-15}. The ICMPA
 is in partnership with
the Daniel Iagolnitzer Foundation (DIF),
France.

\end{document}